\documentclass[]{aa}
\usepackage[varg]{txfonts}
\usepackage{graphicx}

\bibpunct{(}{)}{;}{a}{}{,} 

\begin{document}

  \title{The accretion of migrating giant planets}
  \author{Christoph D\"urmann \and Wilhelm Kley}
  \institute{Institute of Astronomy and Astrophysics, Universit\"at T\"ubingen, Auf der Morgenstelle 10, 72076 T\"ubingen, Germany \\
    \email{christoph.duermann@uni-tuebingen.de}}
  \abstract{
  }{
    Most studies concerning the growth and evolution of massive planets focus either on their accretion or their migration only.
    In this work we study both processes concurrently to investigate how they might mutually affect one another.
  }{
    We modeled a two-dimensional disk with a steady accretion flow onto the central star and embedded a Jupiter mass planet at 5.2\,au.
    The disk is locally isothermal and viscosity is modeled using a constant $\alpha$.
    The planet is held on a fixed orbit for a few hundred orbits to allow the disk to adapt and carve a gap.
    After this period, the planet is released and free to move according to the gravitational interaction with the gas disk.
    The mass accretion onto the planet is modeled by removing a fraction of gas from the inner Hill sphere,
    and the removed mass and momentum can be added to the planet.
  }{
    Our results show that a fast migrating planet is able to accrete more gas than a slower migrating planet.
    Utilizing a tracer fluid we analyzed the origin of the accreted gas  originating predominantly
    from the inner disk for a fast migrating planet.
    In the case of slower migration, the fraction of gas from the outer disk increases.
    We also found that even for very high accretion rates, in some cases gas crosses the planetary gap from the inner to the outer disk.
    Our simulations show that the crossing of gas changes during the migration process as the migration rate slows down.
    Therefore, classical type II migration where the planet migrates with the viscous drift rate and no gas crosses the gap is no general process but may only occur for special parameters and at a certain time during the orbital evolution of the planet.
  }{}
  \maketitle
  \section{Introduction}
    Mass accretion onto giant planets during their growth phase is a delicate process depending on many aspects of the disk structure
    and a variety of physical processes.  One of these is the radial migration of the planet.
    If the planet is migrating through the disk, important parameters like density, temperature, and viscosity change depending on the migration rate.
    This has implications on the gap opened by the planet and might eventually  alter accretion onto the planet.
    The simulations carried out for this paper cover global aspects such as planetary gaps and migration in order to study their influence on the accretion process.

    The main driving force for the migration of massive planets is the interaction with the surrounding protoplanetary disk.
    The different migration regimes and the contributing physical effects are discussed in several recent reviews on planet disk interactions by \citet{kley2012planet}, \citet{baruteau2013recent}, and \citet{baruteau2014planet}.
    The general idea of type II migration was devised by \citet{ward1982tidal} and \citet{lin1986tidal} and states that a planet massive enough ($> 0.5\,M_\mathrm{Jup}$) will open a gap in the gas disk around the protostar.
    Under the assumption that the gap follows the viscous evolution and moves inward with the radial viscous speed, and because the gap edge creates
    torques repelling the planet, it is locked in the middle of the gap and has to follow the viscous disk evolution.

    This picture has to be revised after recent findings showing that a planet in a gap usually migrates at rates faster or slower than the viscous rate \citep{edgar2007giant,morbidelli2007dynamics,duffell2014migration,duermann2015migration}.
    This is important for models of population synthesis because the role of type II migration is not yet clear \citep{hasegawa2013giant}.
    For classical type II migration it is crucial to have a gap which separates the inner from the outer disk because only in this case is the gap forced to follow the viscous evolution of the gas.
    If gas can cross the gap this means there are different channels of transportation for the gas and the viscous speed of the gas does not need to coincide with the migration rate of the planet.
    This cannot be modeled separately from planet migration because for stationary planets in equilibrium there will automatically be a gas flow across the gap matching the disk accretion rate, something also found by \citet{fung2014empty}.
    Also, for low-mass planets it has become clear that to study migration, the analysis of static torques is not sufficient.
    Recent work by \citet{paardekooper2014dynamical} and \citet{pierens2016migration} showed that dynamical torques resulting from the migration must also be considered.

    Giant planet accretion has been investigated by several groups in various ways.
    One kind of research focuses on the inner structure and the atmosphere of the growing planet \citep{pollack1996formation,hubickyj2005accretion}.
    They do not explicitly model the protoplanetary disk around the planet but this way they have a very high resolution of the interior of the planet and its close surroundings.
    Another kind of simulation models the whole disk and inserts the planet at some stage in its evolution. See \citet{lubow1999disk}, for example.
    Because the whole disk, in two-dimensions (2D) or even three-dimensions (3D), has to be modeled, the limiting factor is the resolution in the vicinity of the planet, which can be overcome to some degree by adaptive mesh refinement or nested grids, as used by \citet{dangelo2003orbital}.

    In this paper we focus on the mutual effects of the gas accretion and giant planet migration.
    In particular, we investigate where the accreted gas originates in the disk and study influence of the gap.
    With respect to type II migration we analyze whether gas can cross the gap and how this is affected
    by the mass accretion rate onto the planet.

    In Sect. \ref{setup} the numerical model we used is explained, and
    in Sect. \ref{method} we describe the method we used to model accretion.
    In Sect. \ref{fractions} we discuss the effect of varying the accretion rate and how this affects the migration rate.
    The effects of different models of accretion are discussed in Sect. \ref{accretion-models}. In Sect. \ref{origin} we show results of where the accreted gas comes from and how it can move through the gap.
    We discuss the results in Sect. \ref{conclusion}

  \section{Model setup} \label{setup}
    To model the migration and accretion of embedded planets in disks we use 2D locally isothermal models that are calculated with the hydrodynamical code NIRVANA \citep{ziegler1997nested,ziegler1998nirvana}.
    We use a setup that would, without a planet, represent an accreting equilibrium disk including a mass flow onto the central star
    with constant accretion rate $\dot{m}$.
    The initial and boundary conditions are identical to our simulations in \citet{duermann2015migration}.
    Therefore, the disk surface density is given by
    \begin{equation}
      \Sigma(r) = \frac{\dot{m}}{3\pi\alpha h^2 \sqrt{G M_\odot}}r^{-1/2} = \Sigma_0 r^{-1/2},
      \label{eq:surface_density}
    \end{equation}
    and the initial radial velocity is
    \begin{equation}
      u_r = u_r^\mathrm{visc} = -\frac{3}{2}\alpha h^2 r \Omega_\mathrm{K},
      \label{eq:viscous_flow}
    \end{equation}
    where $\Omega_\mathrm{K}$ is the Keplerian orbital frequency and $u_r^\mathrm{visc}$ is the speed of the radial viscous flow for an equilibrium disk with constant $\dot{m}$.
    The outer boundary conditions are a forced inflow at a constant $\dot{m}$ with the initial radial velocity.
    At the inner boundary we have a forced outflow with the initial radial velocity and zero gradient for the density and energy density.
    The computational domain covers the region between $r_\mathrm{min}=0.3r_0$ and $r_\mathrm{max}=3.0r_0$ with $r_0=5.2\,\mathrm{au}$,
    and $\Sigma_0 = \Sigma(r_0)$.
    At the inner boundary there is a damping region between $r_\mathrm{min}$ and $1.25 r_\mathrm{min}$ where all components of the velocity are damped to the azimuthal average with increasing strength closer to the boundary.
    At the outer boundary the damping region begins at $0.8 r_\mathrm{max}$ and there the $\tau_{r,r}$ element of the stress tensor is increased for the radial velocity update to suppress the viscous overstability \citep{kley1993two}.
    We resolve the domain with 389 radial and 901 azimuthal equally spaced grid cells.
    The planet is introduced on a circular orbit at $a_0 = r_0 = 5.2\,\mathrm{au}$ around a solar mass star.
    The viscosity is given by the $\alpha$-model with $\alpha=0.003$
    and the disk scale height is $h=0.05$.

    In the first 500 orbits, corresponding to approximately 6000 years, the planet is held on its initial circular orbit to allow the disk time to adapt to the presence of a massive planet.
    During this phase, the planet opens a gap in the disk as would happen during a slow growth process in the disk.
    We do not reach full equilibrium in this 6000 years but, because we are interested in migration, the main goal is to make sure the disk is able to adapt to the migrating planet without introducing strong artificial effects as a result of the sudden release.
    Another method we use to reduce the effects of the release of the planet is to slowly switch on the disk torques acting on the planet over
    six orbits.

    The torques generated within $f_\mathrm{t}=0.8R_\mathrm{H}$ are reduced applying a Fermi-like tapering function
    \begin{equation}
      f_\mathrm{taper}(r_\mathrm{cell})=\frac{1}{1+\exp\left(-\frac{r_\mathrm{cell}-f_\mathrm{t}}{0.1f_\mathrm{t}}\right)},
      \label{eq:tapering}
    \end{equation}
    where $r_\mathrm{cell}$ is the distance of the cell to the planet.
    The total torque is then calculated by
    \begin{equation}
      \Gamma_\mathrm{tot} = \sum\limits_\mathrm{cells} f_\mathrm{taper}(r_\mathrm{cell}) \Gamma_\mathrm{cell} \,,
    \end{equation}
    where $\Gamma_\mathrm{cell}$ is the torque exerted by one grid cell on the planet.
    This way gas which is already bound to the planet does not contribute to the migration

  \section{Accretion method} \label{method}
    \begin{figure}
      \centering
      \includegraphics[width=.6\columnwidth]{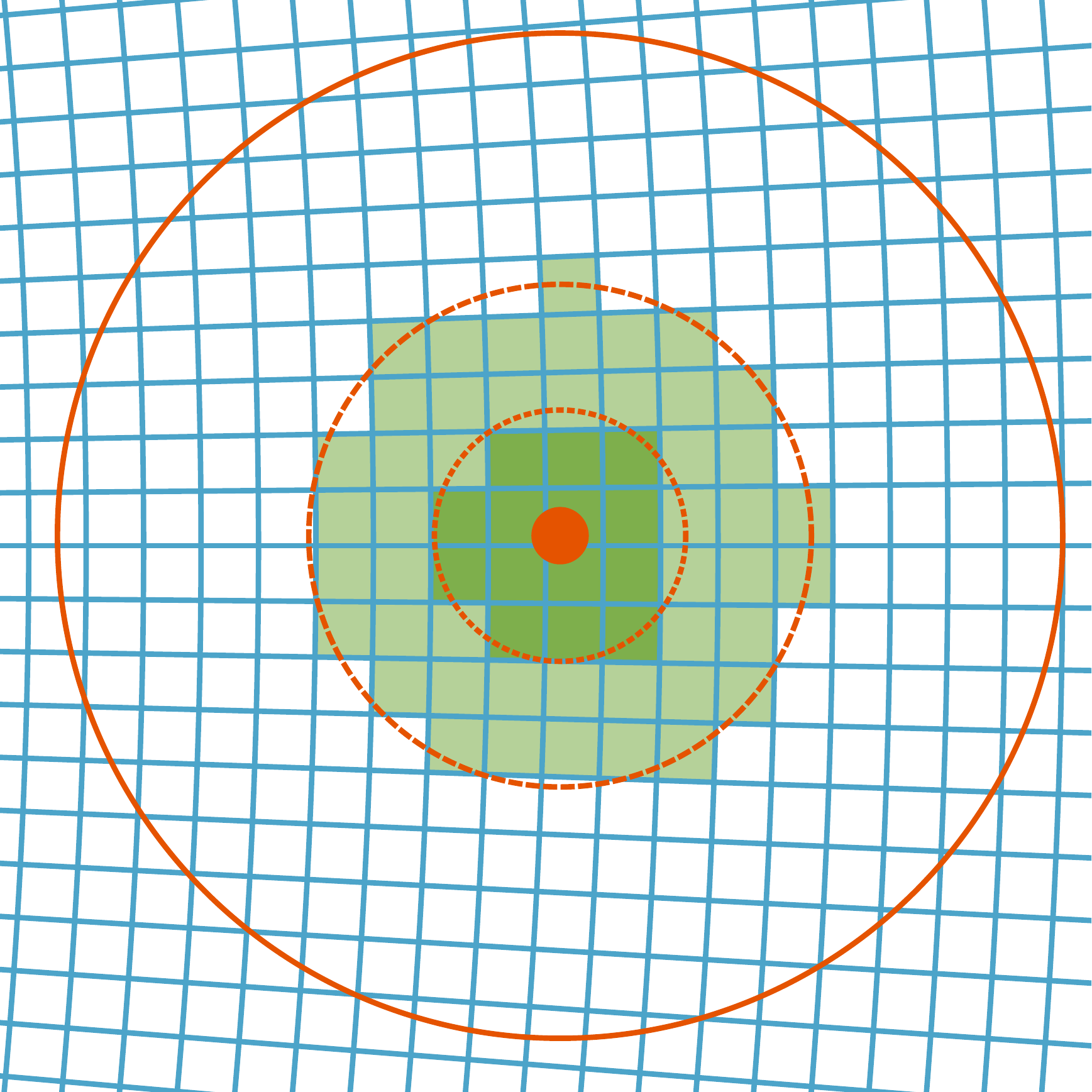}
      \caption{
        Section of the grid around a planet (red dot) and its Hill Radius $R_\mathrm{H}$ shown as a red circle.
        The smaller dashed circles have radii of $R_\mathrm{H}/2$ and $R_\mathrm{H}/4$.
        The accretion mechanism we use removes gas from the lighter and darker shaded cells.
        In the darker shaded cells the accretion fraction is increased by a factor of two.
        Because the planet is allowed to move in the grid it is not centered in a cell or located on cell boundaries.
        \label{fig:accretion-method}
      }
    \end{figure}
    To model the mass accretion onto the planet we rely on the method already used in our previous work \citep{duermann2015migration}, originally described by \citet{kley1999mass}, but with some refinements.
    At each time-step the gas density in cells within half of the Hill radius $R_\mathrm{H}=a_\mathrm{P}\left(M_\mathrm{P}/3M_*\right)^{1/3}$ around the planet is reduced by a fraction $f_\mathrm{acc}{\Delta t}\Omega$ where $a_\mathrm{P}$ is the semimajor axis of the planet,  $\Delta t$ is the time step, $\Omega$ the Keplerian orbital frequency at the initial planetary orbit, and the accretion fraction $f_\mathrm{acc}$ is a free parameter.
        In addition, the accretion in the inner quarter of the Hill radius is twice as high as in the outer quarter, as explained in Fig. \ref{fig:accretion-method}.
    \citet{tanigawa2002gas} showed that the accretion rate converges for a fixed $f_\mathrm{acc}$
if the accretion zone is small enough, $\lesssim 0.07R_\mathrm{H}$, but as we are interested in modeling different accretion rates and our method is not realistic anyhow, we chose a larger region, which is beneficial for a smooth accretion rate with a planet moving through the grid.
    The parameter, $f_\mathrm{acc}$, defines a depletion timescale, at which the affected cells in the Hill sphere will be emptied if there is no flow of gas refilling them.
    From the accretion fraction the depletion timescale can be calculated by $\tau_\mathrm{acc}=(f_\mathrm{acc}\Omega)^{-1}$ which gives the timescale at which the inner part of the Hill sphere would be emptied as $\rho = \rho_0 \exp(-t/\tau_\mathrm{acc})$ if it would not be refilled by surrounding gas.
    The mass removed this way is measured and either added to planet mass or not, depending on the particular model.
    Increasing the planet's mass affects its dynamical mass as well as the gravitational potential.
    In calculations where we include a tracer gas it can be accreted the same way as the gas and the amount of removed tracer is measured separately.

    In addition to the mass accretion, we also take the possibility of momentum accretion into account in our models.
    In the simplest assumption, the specific linear and angular momentum of the planet is not affected by the increased mass, thus the velocity of the planet in the disk is not modified by accretion.
    But the accreted gas, depending on its position in the disk, may have a relative velocity with respect to the planet and therefore this simple assumption does not hold necessarily.
    To account for this, we measure the linear momentum of the removed gas in each cell and can add its sum to the linear momentum of the planet as we increase its mass.
    This way we avoid taking into account the rotation (spin) of the planet around its own axis and ensure conservation of linear momentum.
    Obviously the linear momentum and the orbital angular momentum of the planet are strongly coupled, so if the planet changes its linear momentum this way, it will also change its orbital angular momentum around the star.

  \section{Accretion without growth} \label{fractions}
    \begin{figure}
      \includegraphics[width=\columnwidth]{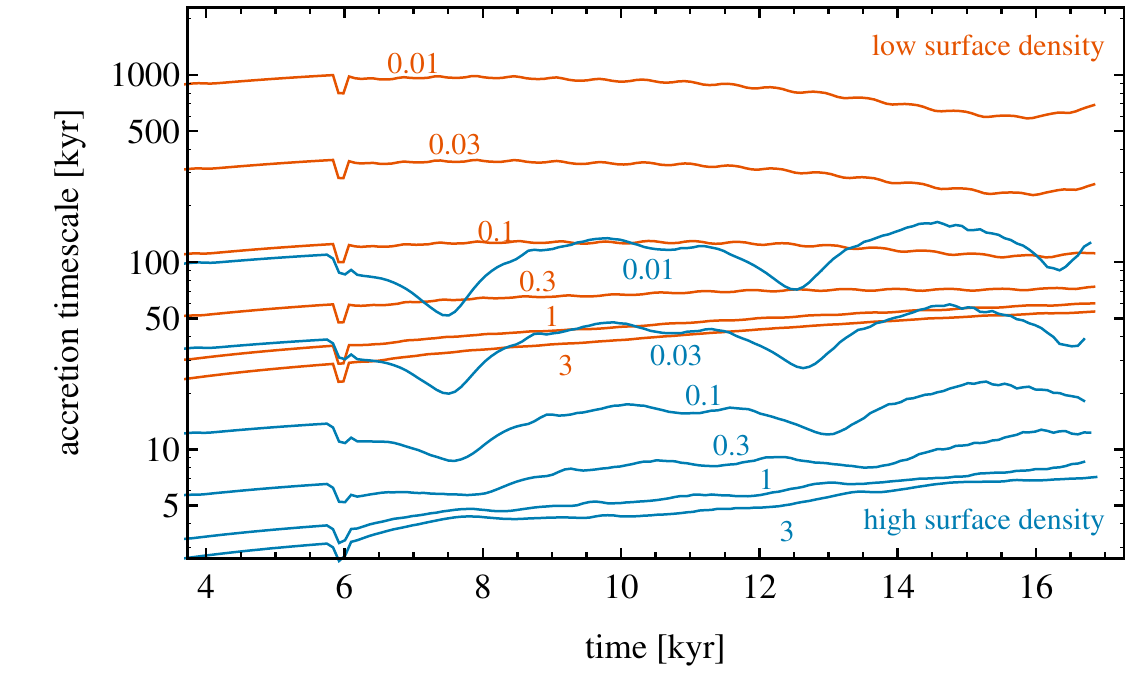}
      \caption{Accretion timescale for Jupiter-mass planets with different accretion fractions $f_\mathrm{acc}$ given as numbers close to the lines.
      For the first 6000 years the planets were on fixed orbits and their mass kept constant, while already gas was removed in the vicinity of the planet.
      Then the planets began to migrate and the gas removal was continued, but it was not added to the mass of the planet.
      The models were calculated with surface density $\Sigma_0=88\,\mathrm{g/cm^2}$ (red) and $\Sigma_0=880\,\mathrm{g/cm^2}$ (blue), respectively.
      \label{fig:mass_doubling_times}
      }
    \end{figure}
    Our model of accretion depends on the choice of an appropriate accretion fraction, $f_{\mathrm acc}$,
   that cannot be derived from physical arguments in our simulations. To model these processes, one would need a very high spatial grid resolution close to the planet and additional physics such as thermodynamics, radiation transport, opacities and even chemistry.
    This is not in the scope of this work and for this reason we chose another approach.
    To find a value of the accretion fraction that leads to accretion rates consistent with previous work we performed a parameter study.
    We place a Jupiter-mass planet in disks with two different surface densities of $88\,\mathrm{g\,cm^{-2}}$ and $880\,\mathrm{g\,cm^{-2}}$ at $5.2\,\mathrm{au}$ corresponding to $\dot{m}=10^{-8}$ and $10^{-7}\,\mathrm{M_\odot\,yr^{-1}}$, respectively.
   For the accretion fractions, $f_\mathrm{acc}$, we chose values between $10^{-2}$ and $3$.
    We only removed the gas in the vicinity of the planet from the simulation but its mass was kept constant in this first series of models
    studied in this section. The effect of a mass change of the planet is studied in the following section.

    For our chosen accretion method a higher disk surface density should lead to a higher accretion rate onto the planet.
    In Fig. \ref{fig:mass_doubling_times} we show the accretion timescale $\tau=m_\mathrm{P}/\dot{m}_\mathrm{P}$ obtained in these calculations.
    In case of low accretion fractions the accretion timescales are directly proportional to the depletion timescale  $\tau \propto \tau_\mathrm{acc} = (f_\mathrm{acc} \Omega)^{-1}$, and the inverse of the disk surface density.
    A factor of ten in the accretion fraction or the surface density reduces the accretion timescale to 10\,\%, hence the blue and red lines are offset initially by a factor of ten.
    For very high accretion fractions this proportionality no longer holds due to the accretion timescale reaching a lower limit because
    in this case the disk is not able to supply any more gas to the planet.
    Because the amount of gas still depends on the disk mass for the different surface densities, the accretion timescale limits are again approximately a factor of ten apart.
    The lower limit of the timescale is approximately 20\,kyr for the low density disk and approximately 3\,kyr for the high density disk, which is in agreement with \citet{lubow1999disk}, \citet{dangelo2002nested} and \citet{machida2010gas}.

    The accretion timescale is not constant with time, but changes during the different phases of the simulations.
    During the first phase of the simulations where the planet is held on a fixed circular orbit the accretion timescale gets larger, meaning slower accretion.
    This can be explained by the introduction of the planet into the undisturbed disk.
    The planet begins to carve a gap and the density near the planet is reduced slowing down the accretion until a new equilibrium is reached.
    After 6000 years the planet is released and allowed to migrate which is the cause of the small perturbations in all the curves at this time.

\begin{figure}
      \includegraphics[width=\columnwidth]{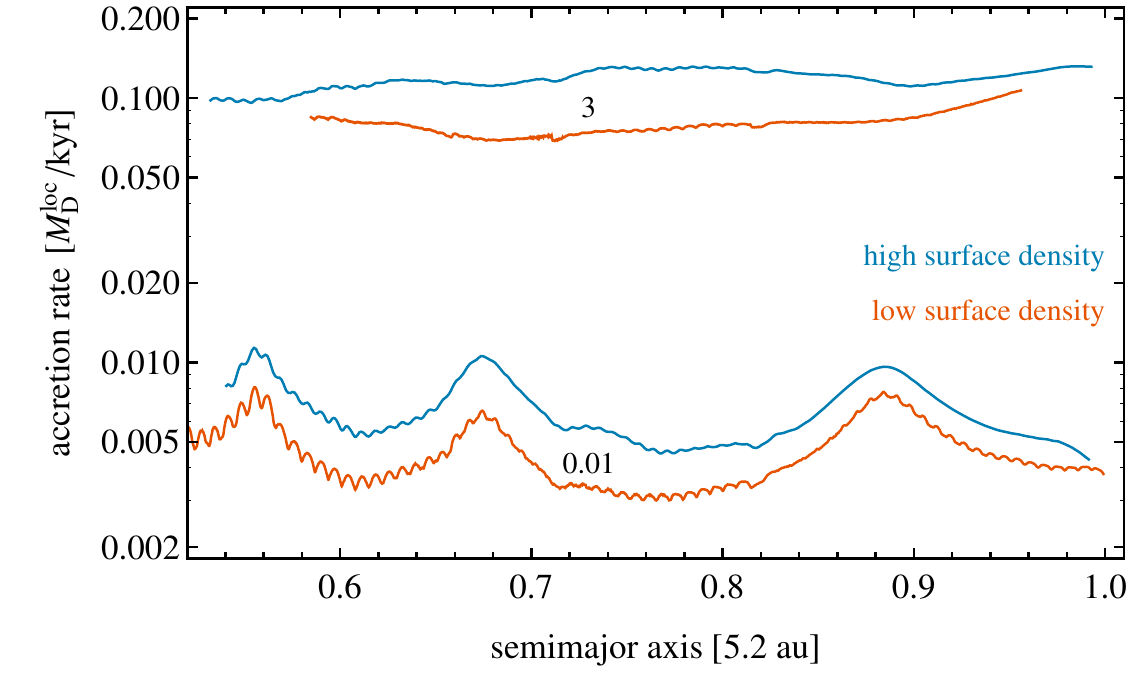}
      \caption{
        Accretion rate for selected models from Fig.~\ref{fig:mass_doubling_times} (measured in units of the local disk mass $M_\mathrm{D}^\mathrm{loc}=a_\mathrm{P}^2 \Sigma_0 a_\mathrm{P}^{-1/2}$ per kyr) as a function of the planet's distance. Only the highest and lowest accretion fractions (numbers) are shown. The red and blue lines correspond to low and high surface density.
        \label{fig:accretion-rate_facc}
      }
    \end{figure}

    In this second phase the accretion timescale does not behave as smoothly as before.
    This is best seen for the calculations with higher disk density (blue lines) as there are visible valleys at 7.5, 12.5 and 16.5\,kyr in Fig.~\ref{fig:mass_doubling_times}, that move to slightly later times for smaller accretion timescales.
    As the planets are migrating with different rates, these valleys appear in all calculations when the planets are at specific positions in the disk, which are $a_\mathrm{P}=0.88, 0.67$ and $0.55\,r_0$, as can be seen clearly in Fig.~\ref{fig:accretion-rate_facc} where the valleys appear as bumps.
    This also holds true for the simulations with lower surface densities, where the migration is much slower and only the valley at $a_\mathrm{P}=0.88$ is visible at approximately 15.5 kyr.
          In Fig. \ref{fig:accretion-rate_facc}, which shows the same models as Fig. \ref{fig:mass_doubling_times}, but for the highest and lowest accretion fraction, the bump at $a_\mathrm{P}=0.67$ is also visible.

\begin{figure}
      \includegraphics[width=\columnwidth]{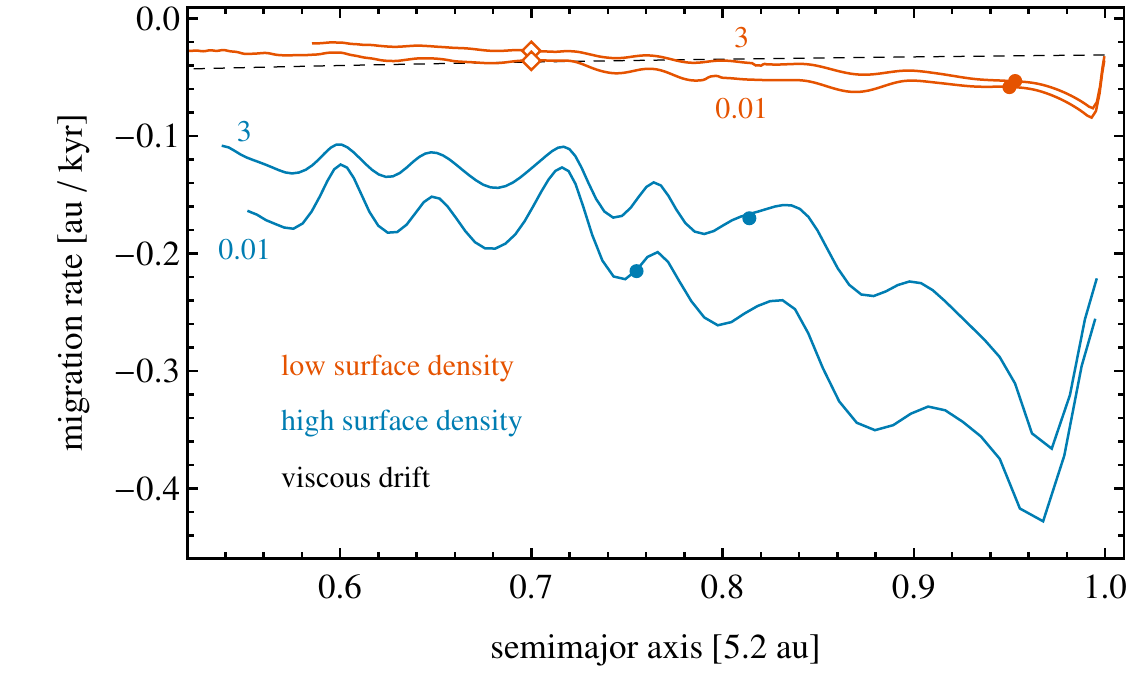}
      \caption{
        Migration rates of the simulations shown in Fig. \ref{fig:accretion-rate_facc}.
        The black dashed line is the radial viscous flow speed for an equilibrium disk with constant $\dot{m}$
       (see Eq. \ref{eq:viscous_flow}), which is independent of the surface density.
        The dots and diamonds mark those times in the evolution which are discussed in section \ref{origin} and Fig. \ref{fig:mdot-profiles}
        \label{fig:migration-rates_accretion-fractions}
      }
    \end{figure}

    To identify the origin of these variations we ran additional simulations where we changed various parameters of the models.
    We found that the location of the bumps dependend on the location of the inner boundary, that is, the value of $r_{\mathrm{min}}$, but not on the value of $r_{\mathrm{max}}$.
    The position does not depend on the numerical resolution
    and is independent of the initial position of the planet.
    However, we found some dependence on the strength of the damping at $r_\mathrm{min}$ because the damping changes the position of the effective inner boundary felt by the planet.
    The position of the bumps also depends on the boundary condition (open or closed inner boundary) and on the sound speed in the disk.
    Here, a higher disk aspect ration $h$ shifts the bumps to smaller radii and vice versa.
    This suggests that the features might be related to sound wave reflections by the inner boundary and their interaction with the migrating planet but future work is needed to investigate this effect in detail.
    The increase of the accretion rate is caused by an increased density in a region very close to the planet ($\lvert\vec{r}-\vec{r_\mathrm{P}}\rvert\la R_\mathrm{H}/2$) while the global structure of the disk is unchanged.
    This also explains why the bumps disappear for higher accretion fractions.
    If the accretion is strong enough, no build-up of gas close to the planet is possible.
    The variation of the accretion timescale is in all cases  a factor of approximately two.
    At the moment we do not fully understand this effect, but it does not influence the overall global evolution of the disk.
    Our additional calculations show no significant variations in the torques (and thus the migration rate) for different values of the damping, the location of the inner boundary (within reasonable limits) or the size of the damping region.
    The migration rates in Fig. \ref{fig:migration-rates_accretion-fractions} show no correlation with the accretion rate in Fig. \ref{fig:accretion-rate_facc}.
    As we are mainly interested in comparing the migration behavior of the planet at different accretion rates, the absolute values of accretion may differ without invalidating our results.


    The accretion rate depending on the planet's position in the disk is displayed in Fig. \ref{fig:accretion-rate_facc}.
    The accretion rate is given in local disk masses $M_\mathrm{D}^\mathrm{loc} = \Sigma_0 a_\mathrm{P}^{3/2}$ per kyr to compare the accretion rate of the different surface densities.
    The main difference remaining is that the planet in the more massive disk (blue line) is migrating much faster as can be seen
    in~Fig. \ref{fig:migration-rates_accretion-fractions}.
    The higher migration rate of the planet leads to a higher accretion rate, because more gas can be supplied to the Hill sphere.
    The accretion rates onto the planet in the high accretion limit exceed the viscous accretion rate of the disk onto the star by a factor of approximately five in the beginning, reducing to a factor of two after 11000 years of evolution.
    We already discussed the fact that planets do not migrate at the equilibrium viscous speed but can be faster or slower  in \citet{duermann2015migration}.

    Accreting planets migrate slower, as shown in Fig. \ref{fig:migration-rates_accretion-fractions}.
    In case of the high accretion rate limit ($f_\mathrm{acc}=3$) it is slowed down by up to 30\,\% compared to the slow accretion with $f_\mathrm{acc}=0.01$ (which does not differ in migration rate from non-accreting simulations) for both high and low surface density.
    In the models discussed in this section the mass of the planet was not increased during the migration.
    Therefore, the changes in the migration rate due to different accretion rates are only an effect of the reduced surface density because of the gas removal from the Hill sphere.
    The lower density reduces the torques and as a result slows down the migration and in a more massive disk this effect will be stronger because more mass can be removed.

  \section{Models with mass growth} \label{accretion-models}
    In addition to the previous models, in order to understand the influence of the planetary mass accretion on the migration of massive planets we compared different methods of accretion. For comparison reasons, in the first model we only removed the gas near the planet and did not add it to the planet (model named R), as in the models discussed in section \ref{fractions}.
    In the following run we added it to the planet mass and kept the specific linear momentum constant (RA) and in the last one we added it and also the linear momentum of the removed gas to the planet (RAM).
    As a reference, we also ran calculations where we did not even remove the gas from the simulation (N).
    To estimate the maximum possible effect, we chose the accretion fraction $f_\mathrm{acc}=3.0,$ which is in the saturated upper limit of the accretion rate and the disk surface density $\Sigma_0=880\,\mathrm{g\,cm^{-2}}$.
    In the first 6000 years (500 orbits) the planets are held on circular orbits and, while the gas is already removed from the simulation, the mass is kept constant at $M_\mathrm{P}= M_\mathrm{Jup}$.
    Then the accretion is switched on (according to the respective model) and the planets move according to their interaction with the gas disk.

    \begin{figure}
      \includegraphics[width=\columnwidth]{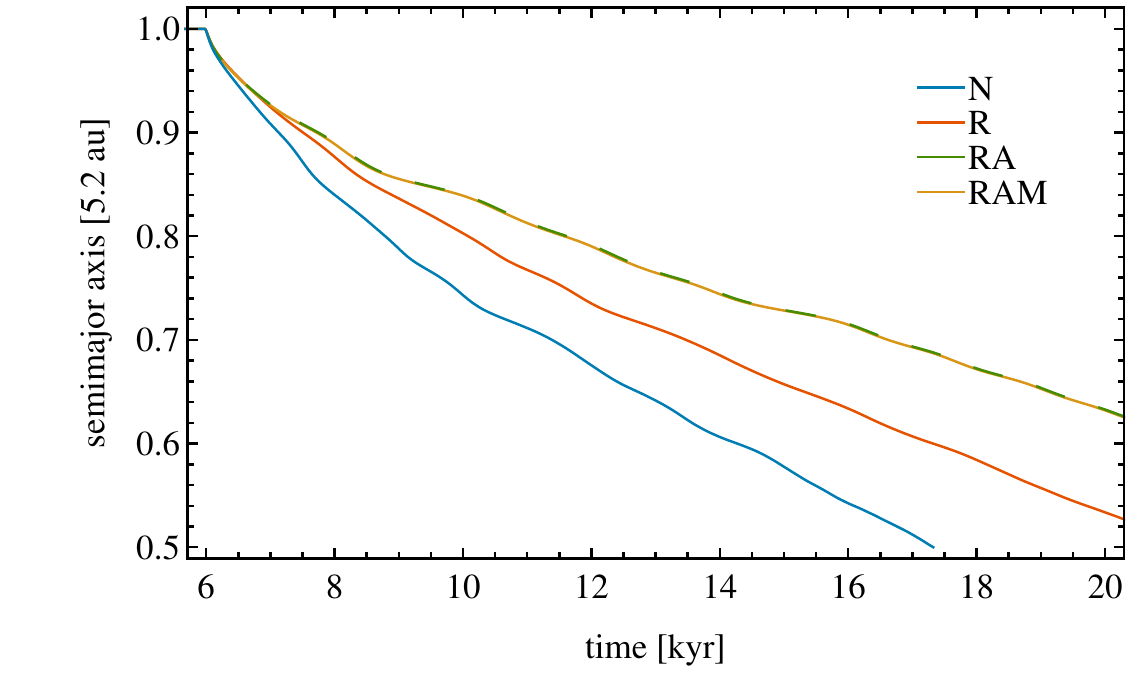}
      \caption{Orbital evolution of an initially Jupiter-mass planet. In model R the gas is only removed, but not added to the planet's mass.
      In model RA the gas is accreted and increases the planet's mass and in RAM  the linear momentum of the accreted gas is also added to the planet.
      Model N is for comparison, no gas is removed in the vicinity of the planet.
      Model RA and RAM are very similar so they are plotted as dashed lines.
      \label{fig:different_accretion_models}
      }
    \end{figure}
    \begin{figure}
      \includegraphics[width=\columnwidth]{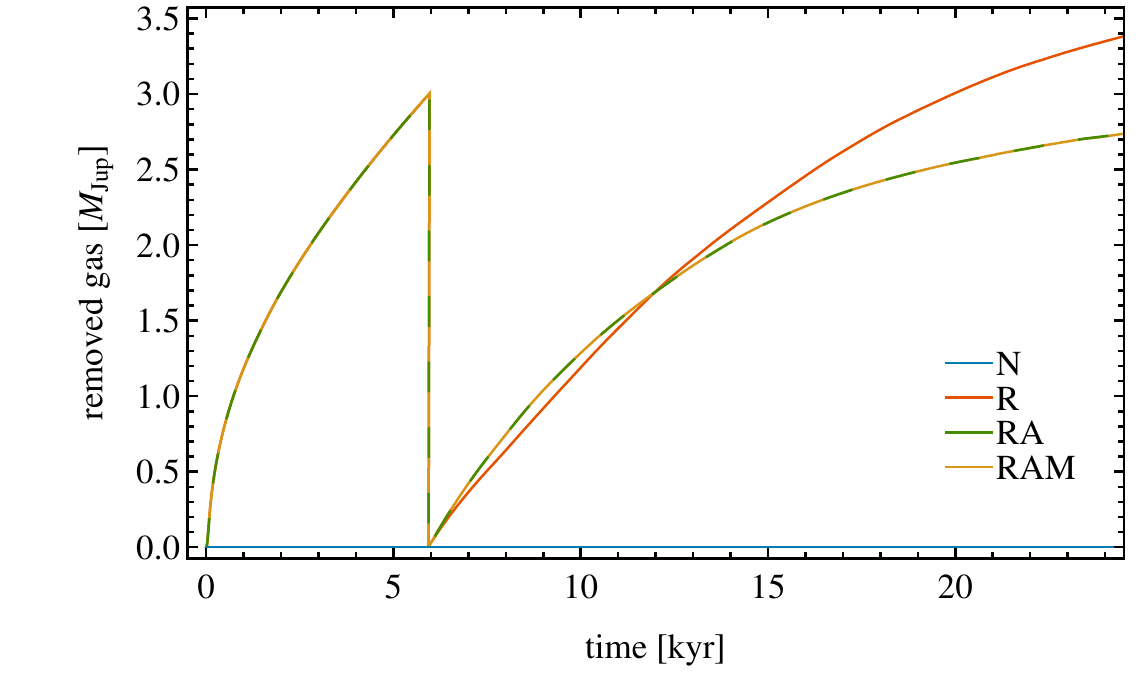}
      \caption{Removed gas for the different models considered in Fig. \ref{fig:different_accretion_models}.
      Before the migration is switched on at 6000 years the planets do not increase their mass and behave like model R in order to start the evolution with $1\,M_\mathrm{Jup}$.
      After the release, we reset the counter and in models RA and RAM the mass is now increased according to the removed gas.
      Again the line of model RA is shown as a dashed line to show model RAM behind.
      \label{fig:different_accretion_models_mass}
      }
    \end{figure}
    The evolution of the semimajor axis in the different models is shown in Fig. \ref{fig:different_accretion_models}.
    Directly after the release of the planets, the disk has still to adapt to the now moving planet.
    This transition is smoothed by gradually switching on the torques acting onto the planet over six orbits.
    After a few 100 years the disk has adapted to the moving planet, and the planet and the disk bear no history of the migration to that point, which we tested by releasing the planet from circular orbits with different initial radii.

    \begin{figure}
      \includegraphics[width=\columnwidth]{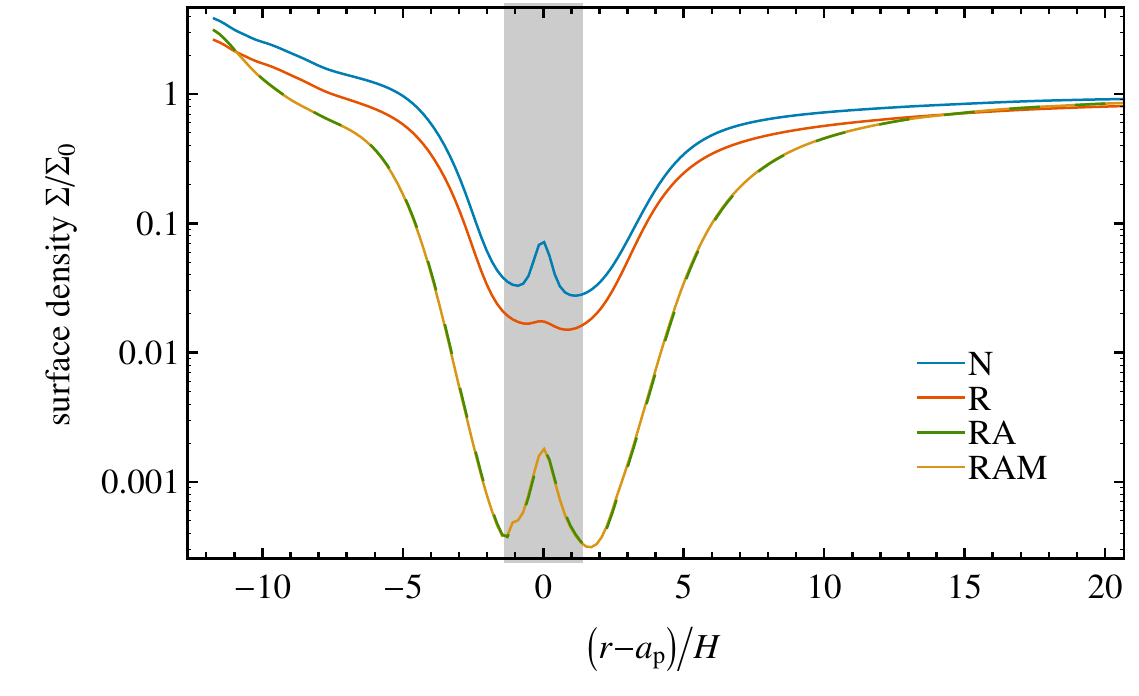}
      \caption{
        Surface density profile close to the planet for the different models discussed in Sec. \ref{accretion-models} when the planet has reached $a_\mathrm{p}=0.75r_0$.
        The distance to the planet is given in units of the local disk scaleheight.
        The shaded region shows the radial extent of the Hill radius.
      \label{fig:gap-profile-compared}
      }
    \end{figure}
    \begin{figure}
      \includegraphics[width=\columnwidth]{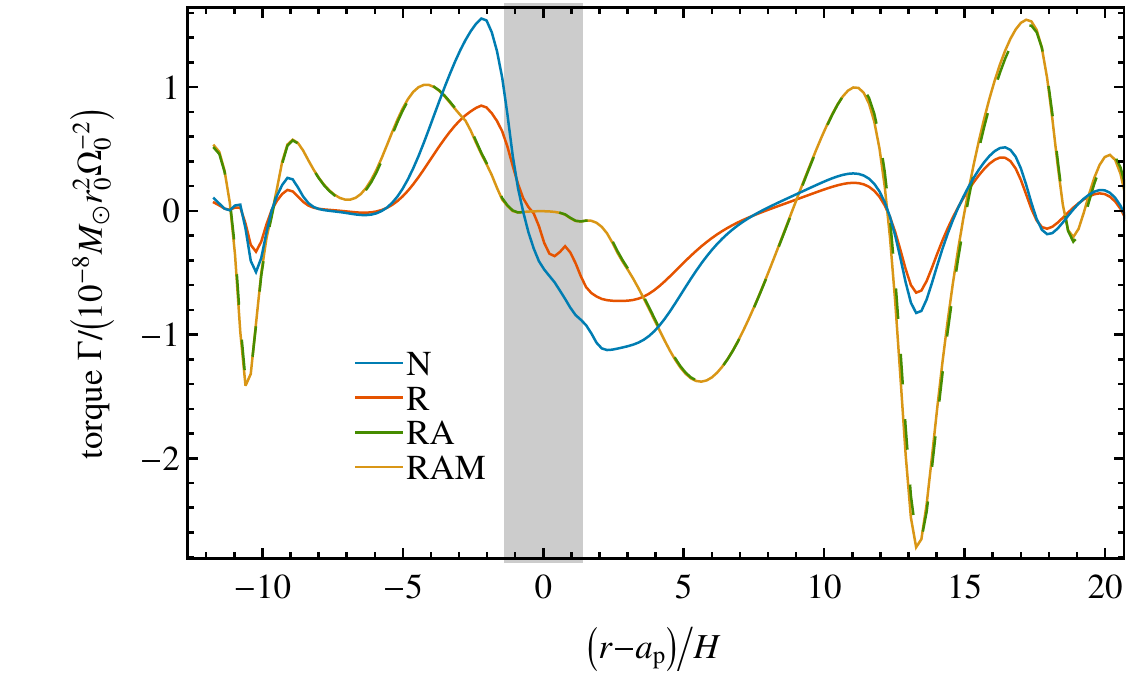}
      \caption{
        Torque profile close to the planet similar to Fig. \ref{fig:gap-profile-compared}.
        Note that the excluded torques (see Eq. \ref{eq:tapering}) are not shown in this graph.
        Torques inside the shaded region are from outside the Hill sphere (e.g., the horseshoe region).
      \label{fig:torque-profile-compared}
      }
    \end{figure}

    In Fig. \ref{fig:different_accretion_models_mass} we show the amount of gas removed in the vicinity of the planet.
    During the first phase where the planet is on a fixed circular orbit the accretion is very high in the beginning as the slope is nearly vertical.
    At this time the planet begins to carve a gap in the disk and the density in its vicinity is still high.
    As soon as the gap develops, the slope decreases and the accretion becomes slower.
    The mass of the planet is not increased until the planet is free to migrate, but in Fig. \ref{fig:different_accretion_models_mass} we show the amount of removed gas up to that time.
    The higher amount of removed gas for the models where the mass of the planet is increased is a result of the higher mass of the planet.
    Since the Hill sphere is growing with the planet's mass, the accretion is enhanced in the beginning, but as the gap is deepened it reduces the amount of material close to the planet and accretion is slowed down again.
    After 25 kyr the planets reached a mass of $2.8\,M_\mathrm{Jup}$ (RA) and $3.4\,M_\mathrm{Jup}$ (R) which is in agreement with results by \citet{nelson2000migration}.

    The migration rates of the different models vary significantly.
    Fastest is the migration without gas removal or accretion (model N).
    We show the surface density close to the planet in Fig. \ref{fig:gap-profile-compared} at times where the planets have reached $a_\mathrm{p}=0.75r_0$ (see Fig. \ref{fig:different_accretion_models}).
    Because no gas is removed, the overall surface density is increased compared to the model R and also gas accumulates close to the planet.
    For models RA and RAM, due to the more massive planet, the density in the gap is less than 10\,\% of that in model R.
    Similarly, we show the radial torque profile in Fig. \ref{fig:torque-profile-compared}.
    Models N and R significantly differ only within $5H$ of the planet, resulting from the higher surface density.
    This means model R has to migrate slower than model N.
    Models RA and RAM show identical torques and differ from the other models because the planet at this time has nearly tripled its mass and thus produces stronger torques.
    Because the more massive planet at the same time needs stronger torques to be moved, but these torques do not increase as much because the gap becomes deeper, the net effect is a slower migration.

    The models RA and RAM are indistinguishable by eye in all these calculations.
    When a planet is accreting gas it also has to accrete momentum, otherwise the increased mass at constant momentum would lead to a decreasing velocity, which of course is unphysical.
    The main contribution to the accreted momentum is the orbital momentum, which is very similar for the planet and the gas because they are both nearly on Keplerian orbits.
    However, there are also deviations from this mean motion such as in- or outflow, as in the horseshoe orbits.
    These deviations are captured by the momentum accretion in the RAM models while the RA models handle only the mean motion.
    \begin{figure}
      \includegraphics[width=\columnwidth]{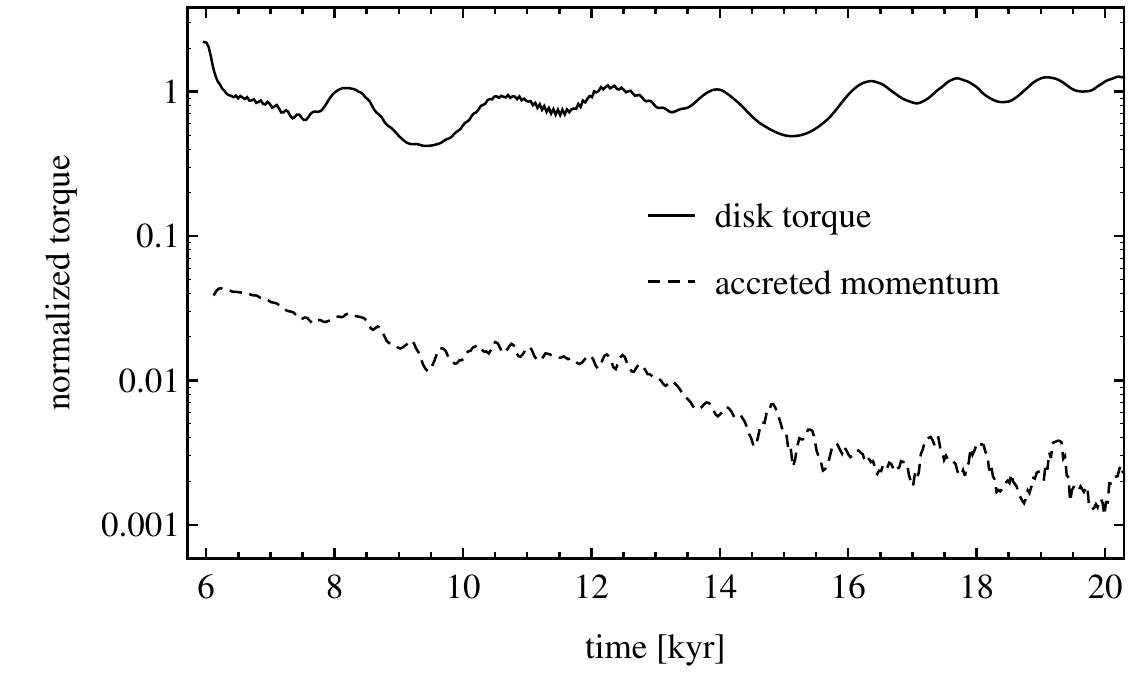}
      \caption{
      Total disk torque in the RAM model from the disk (solid line) and the accreted momentum (dashed line) acting on the planet, normalized to its average.
      The line for the accreted momentum shows only the deviations from the orbital momentum because the orbital momentum does not influence the migration of the planet.
      \label{fig:torque_RAM}
      }
    \end{figure}
    In Fig. \ref{fig:torque_RAM} we show the total torque acting on the planet (disk forces \& accreted momentum) as well as the torque resulting from accreted momentum which is not orbital momentum.
    The contribution of the momentum accretion to the total torque is less than $5$\,\% at the beginning and at later times very small (less than $0.5$\,\%) and therefore negligible.
    Because we cannot resolve the circumplanetary disk, our models do not consider the angular momentum of the removed gas with reference to the planet to analyze its rotation.

    During the evolution, the eccentricity of the planetary orbit does not increase significantly and is below $0.006$ at all times.
    We also monitored the eccentricity of the disk which was well below 0.015 during the calculations.
    For all simulations shown in this paper, the gap edges were steady and displayed no oscillations or eccentricities.


  \section{Origin of accreted gas} \label{origin}
    To investigate the origin of the accreted gas, we implemented a tracer fluid, which is advected with the gas in the simulation.
    It can therefore be removed from the simulation as the accreted gas and the total amount of removed tracer fluid is measured.
    After 10000 years, when the disk had ample time to adapt to the planet and its accretion and migration, we introduced the tracer fluid.
    These times in the orbital evolution are marked in Fig. \ref{fig:migration-rates_accretion-fractions} by round dots.
    The tracer has the same distribution as the gas, but is only present inside the planet's current orbit.
    By measuring the accreted tracer and comparing it to the accreted gas we can determine whether the accreted material comes from the area inside the planet's orbit or from the area outside the planet's orbit.
    Because the planet is migrating during the simulation and therefore inner and outer parts of the disk are mixing and changing with time, we restrict this kind of analysis to 1000 years after the introduction of the tracer gas.
    To account for different migration rates, we used two different disk surface densities of $88$ and $880\,\mathrm{g\,cm^{-2}}$ , and in both cases we considered accretion near the high limit found in section \ref{fractions} with $f_\mathrm{acc}=3$ and in the non saturated case with $f_\mathrm{acc}=0.01$.
    \begin{figure}
      \includegraphics[width=\columnwidth]{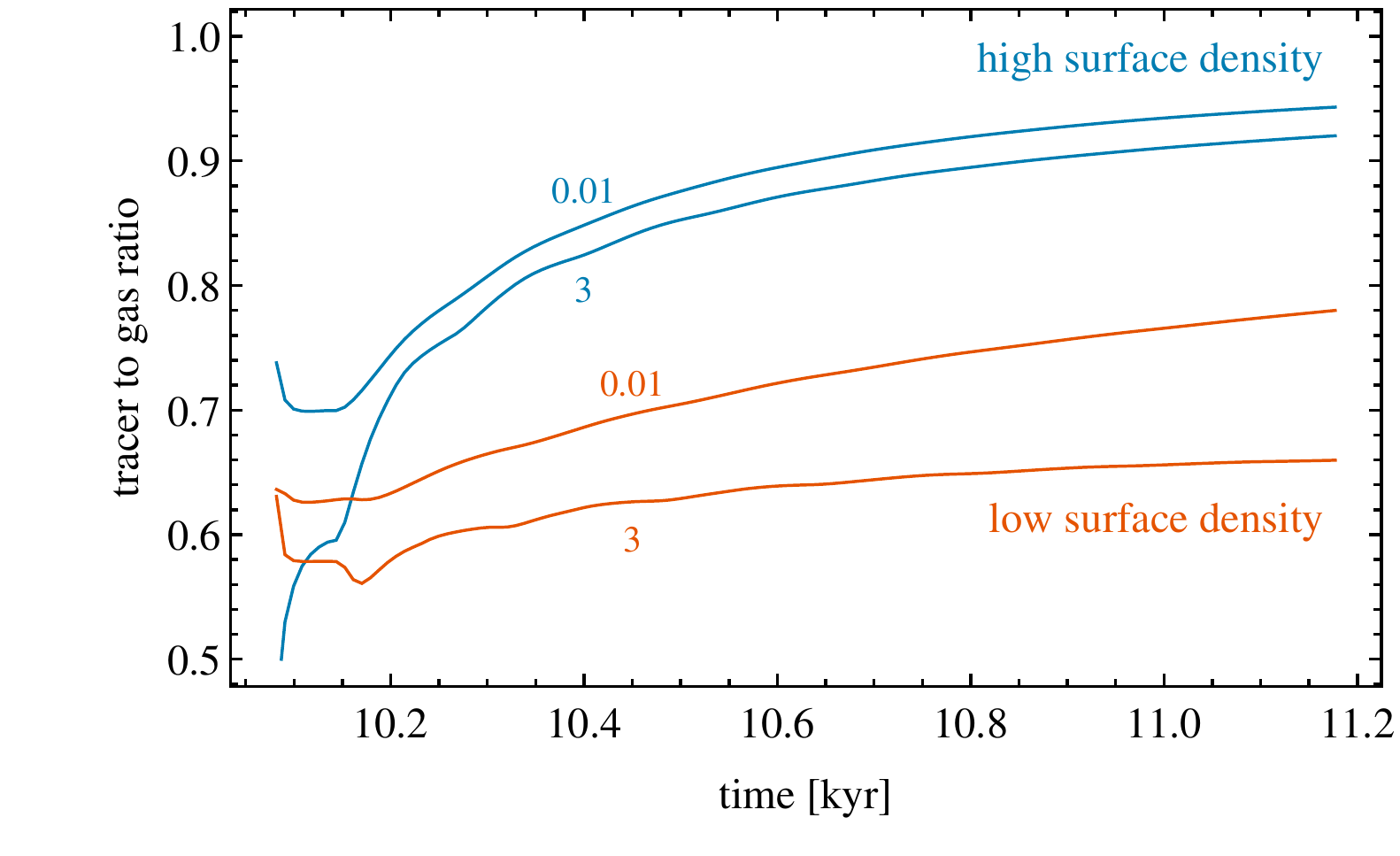}
      \caption{Fraction of the accreted tracer to the accreted gas $m_\mathrm{acc}^\mathrm{tracer}/m_\mathrm{acc}^\mathrm{gas}$.
      The different colors correspond to high and low accretion rates and high and low disk density.
      \label{fig:tracer-fraction}
      }
    \end{figure}

    The ratio $m_\mathrm{acc}^\mathrm{tracer}/m_\mathrm{acc}^\mathrm{gas}$ is shown in Fig. \ref{fig:tracer-fraction} where $m_\mathrm{acc}^\mathrm{gas}$ and $m_\mathrm{acc}^\mathrm{tracer}$ are the amount of gas and tracer accreted by the planet after the tracer introduction.
    The ratio is between $0.5$ and $0.75$ in the beginning but then increases.
    The initial increase is due to the initial distribution which is not in equilibrium because of the flow pattern in the gas, as is the case for horseshoe orbits.
    In case of the more massive disk, which also implies a faster migration rate after only 300 years, more than 80\,\% of the accreted material is originating from the inner disk.
    In the end, the ratio for both accretion fractions is well above 90\,\%.
    Even for the low disk density, and the, therefore, slower migrating planets, the ratio is almost always above 60\,\%.
    The planets in simulations with higher surface density are migrating much faster as can be seen in Fig. \ref{fig:migration-rates_accretion-fractions}.
    The planets in the simulations with higher surface density both migrate approximately $0.03r_0$ while in case of the lower surface density they move only $0.01r_0$ during the 1000 years with the tracer.
    Because they traveled much less than the gap width, they have not migrated far into the initially inner disk.
    This means the planet is still at the edge of the tracer fluid and the ratio of accreted tracer to gas is not only a result of the planets migration deep into the region with the tracer fluid.
    In case of the lower accretion rate more material comes from the inner disk than for the higher accretion rate.
    To understand this behavior, we analyzed the density distribution of the tracer material in the vicinity of the planet.

    \begin{figure*}
      \includegraphics[width=180mm]{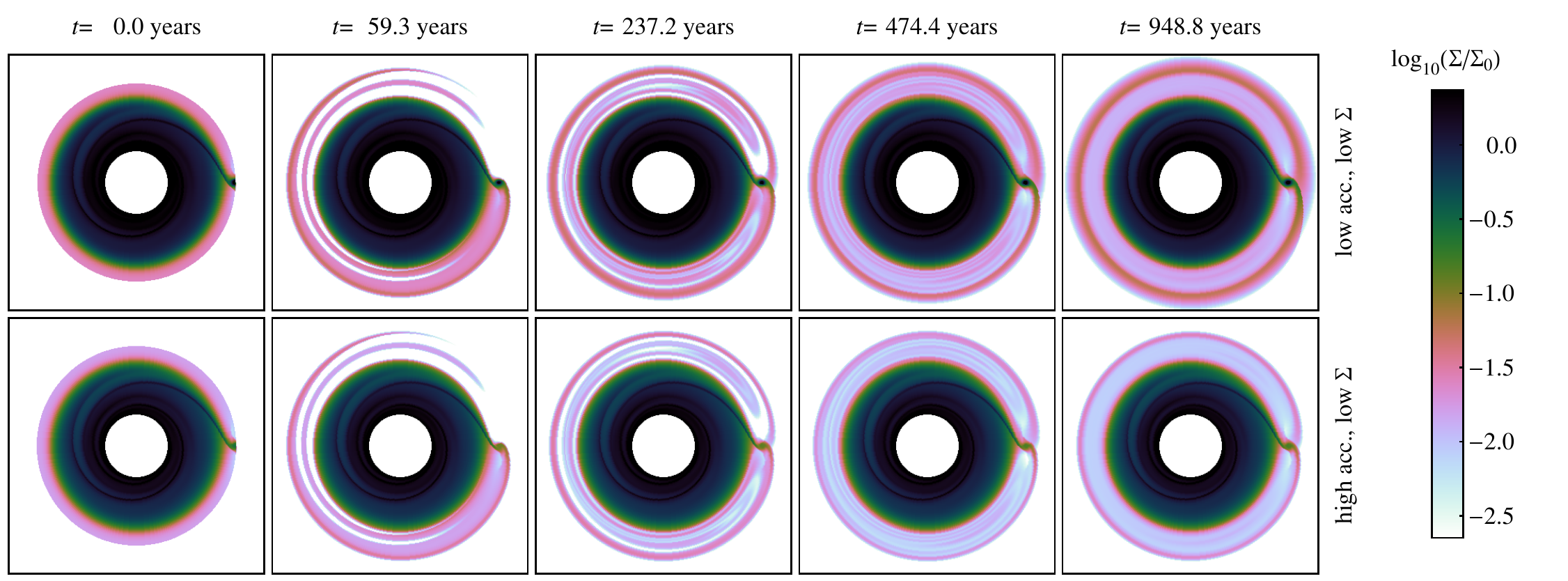}
      \caption{
      Tracer density at different times after the tracer was introduced in the simulation. Both rows show the low surface density ($88\,\mathrm{g\,cm^{-2}}$) with low accretion fraction $f_\mathrm{acc}=0.01$ in the upper and $f_\mathrm{acc}=3.0$ in the lower row. These models correspond to the low surface density models in Fig. \ref{fig:tracer-fraction}.
      \label{fig:tracer-density-timeseries}
      }
    \end{figure*}

    In Fig. \ref{fig:tracer-density-timeseries} the evolution of the tracer with time is shown for the low surface density models in Fig. \ref{fig:tracer-fraction}.
    While for the low accretion rate (upper row) gas crosses the gap and accumulates outside the planets orbit, in case of high accretion of the planet, the tracer is only transported in the horseshoe region (lower row).
    After 240 years the outermost horseshoe orbit is already completely populated with tracer material and with increasing time it only becomes more homogeneous, while for the low accretion rate the region containing tracer is increasing.

    \begin{figure*}
      \includegraphics[width=180mm]{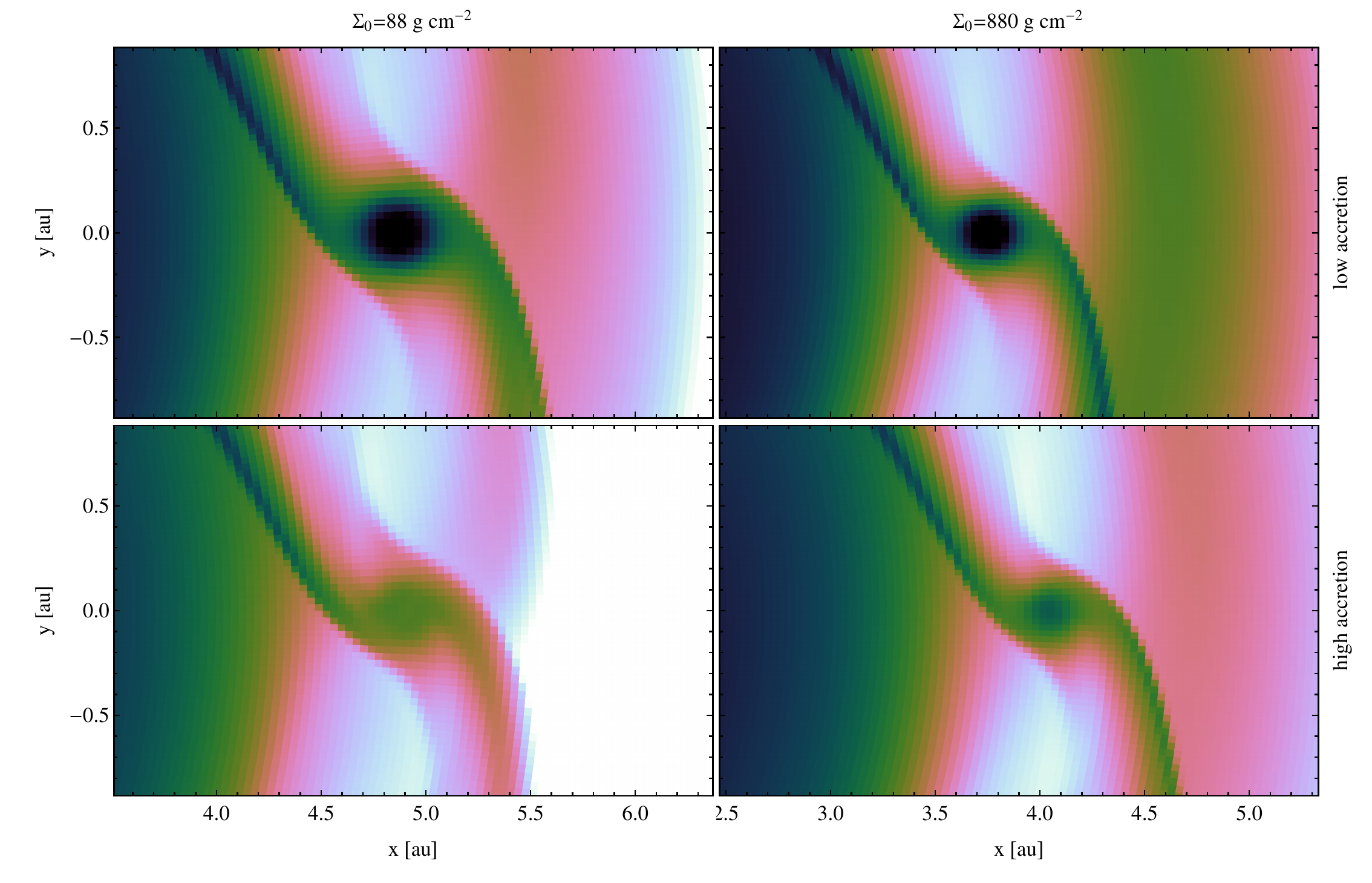}
      \caption{Tracer fluid density near the planet 950 years after the tracer introduction to the inner disk.
      The four panels show the density distribution for low ($88\,\mathrm{g\,cm^{-2}}$) and high ($880\,\mathrm{g\,cm^{-2}}$) disk density from left to right and low ($f_\mathrm{acc}=0.01$) and high ($f_\mathrm{acc}=3$) accretion rate from top to bottom. The color map is the same as in Fig. \ref{fig:tracer-density-timeseries}.
      \label{fig:tracer-density}
      }
    \end{figure*}

\begin{figure}
      \includegraphics[width=\columnwidth]{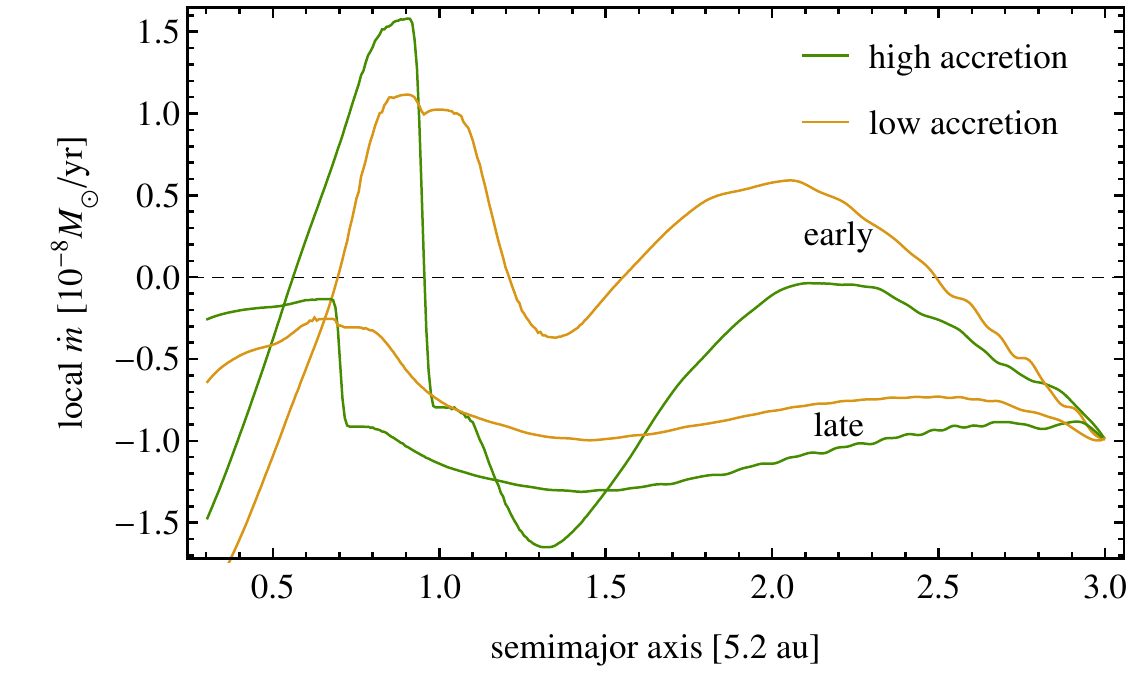}
      \caption{
      $\dot{m}$-profile of the disk (negative values mean inward transport) for several models of the low density case and thus with prescribed disk accretion rate $-10^{-8}\,M_\odot/\mathrm{yr}$ (see. Eq. \ref{eq:surface_density}).
      The profiles are averaged over 1000 years, because they are very sensitive to small disturbances.
      The different colors correspond to the models with high (green) and low (yellow) accretion fractions at early ($t=10$\,kyr) and late times ($t=44$\,kyr) in the evolution.
      The times are marked with red dots and diamonds in Fig. \ref{fig:migration-rates_accretion-fractions}.
      The planets early in the evolution are migrating much faster than the viscous speed,
      in the later models they migrate slower than the viscous speed.
      The jumps at the planet position correspond to the accretion onto the planet, which is therefore much higher in the green lines.
      \label{fig:mdot-profiles}
      }
    \end{figure}

    In Fig. \ref{fig:tracer-density} we show the surface density of the tracer zoomed in to the vicinity of the planet 950 years after the tracer in the inner disk was introduced.
    In addition to the models in Fig. \ref{fig:tracer-density-timeseries} the calculations with high surface density are also shown.
    In the models with low accretion in the top row a pile-up of gas at the position of the planet is clearly visible, but in case of high accretion it nearly completely vanishes.
    The tracer in all cases shows the spiral structure generated by the planet into the outer disk.
    The models in the right column have a higher surface density, thus the planets migrate faster, and because the migration is faster than the viscous radial speed of the gas in all four models at this time (see Fig. \ref{fig:migration-rates_accretion-fractions}), the material passes the gap from the inner to the outer disk and accumulates outside the gap.
    In the simulations with the lower surface density, the migration is still faster than the radial viscous speed but slower compared to the higher density models and hence the transport of material from the inner to the outer disk is slowed down.
    Some transport of material from the inner to the outer disk will of course occur in any case, and even without migration because it is transported through the horseshoe region as shown in Fig. \ref{fig:tracer-density-timeseries}.
    This mechanism can also be seen in the lower left panel of Fig. \ref{fig:tracer-density} where some gas has accumulated in horseshoe orbits and follows their bending near the planet, but there is no pile up of gas outside the horseshoe region.
    This means all the material from the inner disk is accreted onto the planet and also explains why the ratio of accreted tracer to gas in Fig. \ref{fig:tracer-fraction} for that case is reduced compared to the lower accretion rate.
    The material from the inner disk alone is not enough to refill the Hill sphere so there has also to be a flow of gas from the outer disk.
    The same happens, but to a much lesser extent, for the higher disk density where the ratio in the case of the higher accretion fraction is also decreased.
    This can be better understood by looking at the disks local $\dot{m}$ profile shown in Fig. \ref{fig:mdot-profiles}.


    The $\dot{m}$ measures the mass change in the disk by radial movement of gas, therefore a negative $\dot{m}(r)$ means mass from that annulus in the disk is moving inwards, and vice versa with a positive $\dot{m}(r)$.
    The $\dot{m}=-10^{-8}\,M_\odot/\mathrm{yr}$ at the outer domain boundary is a parameter of the simulation and corresponds to the specified value used in Eq. (\ref{eq:surface_density}).
    It would also be the global $\dot{m}$ of the disk in equilibrium without a planet or a non-accreting planet on a fixed circular orbit.
    This was shown in \citet{duermann2015migration} (their Fig. 4), where one can see that reaching the equilibrium takes
   very long even without the planet migrating.
    There we also found the initial outward motion of the disk material in outer parts of the disk as seen here in Fig. \ref{fig:mdot-profiles}.
    These only disappeared after very long times ($>3600$ orbits or 43\,kyr) even for a fixed, non-migrating planet,
   which is consistent with these features disappearing for the late models in our new simulations as presented here.
    For the models with high accretion rate (green) there is a jump at the position of the planet,
     because the planet is removing gas at that position.
    When $\dot{m}$ is always negative close to the planet (as in the late models) this means that,
   although the planet is accreting and migrating, gas is crossing from the outer disk to the inner disk, and
vice versa for the early model with low accretion (yellow), where the gas close to the planet is moving outwards through the gap.
    Also clearly visible is that a fast migrating planet can have a strong impact on the local value of $\dot{m}$ in the whole disk and even induce outward drift of the gas far away from its position.
    The important part explaining the difference in the fraction of accreted material from the inner disk depending on the accretion rate can be seen immediately outside (right) of the planets position.
    For both early and late times, the local $\dot{m}$ for the high accretion models is significantly lower than for the low accretion models and negative meaning a gas flow directed toward the planet.
    Because the planet wants to accrete more gas, the Hill sphere has to refill, and although in most models gas is crossing the gap inside out,
    the inward gas flow at the outer gap edge is stronger, leading to a higher fraction of gas from the outer disk.

    \subsection{Relation to type II migration}
    Our simulations show that gas can cross the gap carved by a giant planet and does so, for example, in cases
    where the planet is migrating faster than the viscous radial speed.
    If the planet is migrating inward faster than the disk, most material will originate from the inner disk.
    Only for high accretion rates and, at the same time, sufficiently low disk surface densities, will the planet be able to accrete all of the gas transported through the gap, as seen in the lower left panel of Fig. \ref{fig:tracer-density}.
    Only in these special cases does the gap impose a barrier between the inner and outer gap and the planet would be in the classical type II regime.
    Also in these cases, however, the migration rate can differ from the classical type II migration rate because this is given
    by the global equilibrium value, which can differ from the actual local value, as for the example seen in Fig.~\ref{fig:mdot-profiles}.
    \begin{figure}
      \includegraphics[width=\columnwidth]{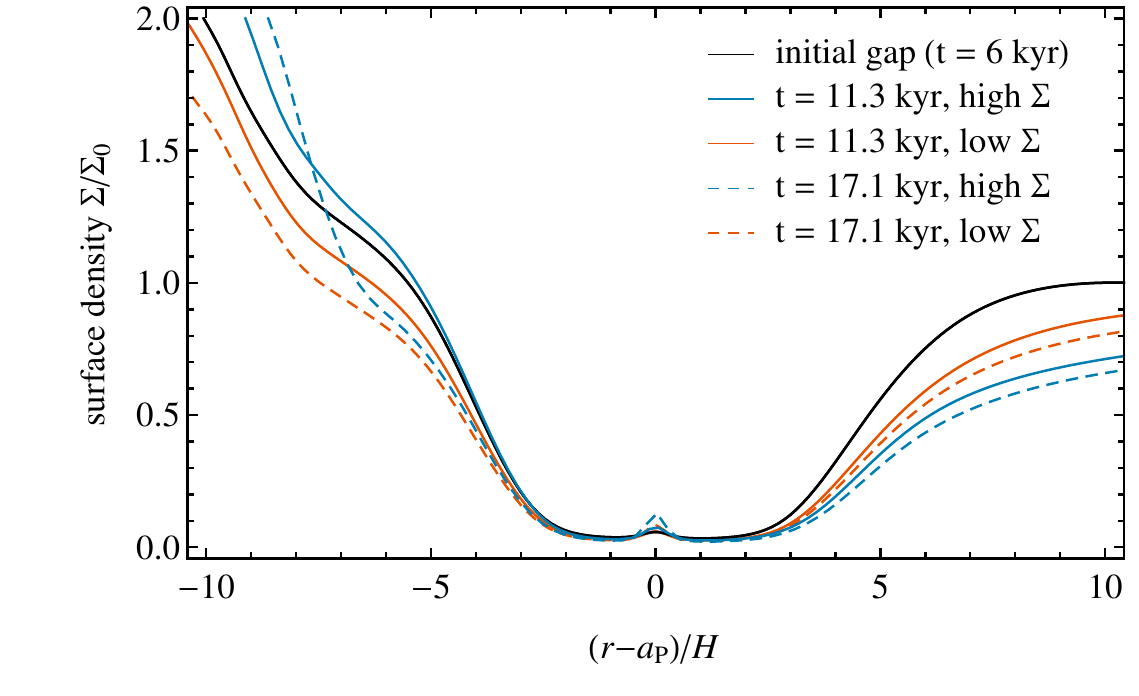}
      \caption{
        Normalized surface density profiles for planets with low accretion rate ($f_\mathrm{acc}=0.01$) at different times during their orbital evolution.
        The horizontal axis is normalized to the planet's position and the disk scale height.
        The black line corresponds to the surface density when the planets are released.
        \label{fig:gap-profiles}
      }
    \end{figure}

    It seems plausible that phases of type II regime can be reached by a migrating planet for a wider range of surface densities
   and accretion rates, but only as a transient state.
    If the planet is migrating inwards faster than $u_r^\mathrm{visc}$ and is accreting a considerable amount of the gas crossing the gap inside out, the surface density at the outer gap edge will be slightly reduced which is shown in Fig. \ref{fig:gap-profiles}.
    Because the outer gap edge exerts a negative torque leading to inward migration, a lower density in this region will slow down migration.
    But at the time when the migration rate equals $u_r^\mathrm{visc}$ the outer gap edge is still depleted and the migration rate will continue to slow down.
    This means, although the radial viscous speed is somehow a natural migration rate, a massive gap opening planet will usually migrate faster or slower if it did not happen to begin its evolution in already perfect equilibrium.
    Due to the fact that, during the overall evolution of planets, the migration transits from fast type I migration or even faster type III migration,
    this perfect equilibrium seems highly unlikely.

  \section{Conclusions} \label{conclusion}
    Our simulations show that the disk sets a limit on how much gas can be supplied to the Hill sphere and thus can be accreted by the planet.
    The upper limit of the accretion rate onto the planet is higher than the accretion rate through the disk onto the star
    by a factor of a few, and reduces with time, which agrees with findings
    of \citet{kley1999mass} and \citet{tanigawa2015final}.

    We show that there are some reciprocal effects between accretion and migration.
    Migration is fastest in the unphysical case where no gas is removed.
    In this case gas is accumulating near the planet and generating large torques because of the small distance to the planet.
    Removal of gas close to the planet reduces these torques and slows down the migration.
    When gas is removed and the planet increases its mass, the migration slows down even more.
    Not only is the gas density close to the planet reduced because of the removal of gas but also because the now more massive planet creates a deeper gap.
    In addition, a growing planet needs stronger torques to keep up its migration rate, but the torque created in the gap region becomes smaller together with the reduced density.
    Also, as a result of the deeper gap, the accretion rate can drop below the accretion rate of a much lighter planet, which does not increase its mass.
    We not only increased the mass of the planet, but also measured the momentum of the removed gas and added it to the planet.
    The simulation clearly shows that this accreted momentum does not have a noticeable effect on the migration of the planet, as the torque it generates is always smaller by a factor of 100 compared to the disk torques responsible for the migration.

    The models with tracer particles to investigate the origin of accreted material gave multiple interesting results.
    For a planet migrating faster than the radial viscous speed (i.e., faster than type II migration), most of the accreted gas originates from the inner disk and this fraction increases when the planet is migrating faster.
    In this case it does not push the inner disk, but gas crosses the gap from the inner to the outer disk.
    Only in cases of high accretion rates and low enough surface densities is the planet able to accrete all of this material.
    In cases where the planet is migrating slower than the viscous speed gas crosses the gap from the outer to the inner disk.
    In both cases gas is able to cross the gap, so it is not separating the inner and the outer disk.
    The rate at which gas crosses the gap is determined by the migration rate of the planet.
    It can be, and often is, faster than the radial viscous speed of the disk which is the local type II migration rate in the disk.

    When the planet is migrating at exactly the viscous speed, and is therefore in the classical type II migration, this is only a transient state during its orbital evolution.
    As we show in Fig. \ref{fig:migration-rates_accretion-fractions} the migration rate is higher in the beginning
    but then drops and reaches the viscous drift rate.
    However, the migration rate is dropping further and becomes even lower than the viscous speed.
    Also, whether or not gas crosses the gap depends not only on the model parameters but also on changes with time, as can be seen in Fig.~\ref{fig:mdot-profiles}.
    The actual migration rate is defined by the state of the disk on both sides of the gap whose evolution depends on a local viscous timescale, which can be different from the global viscous timescale.

    To find the relations that describe the actual migration rate for a given system, further work is needed.
    For the whole picture it is very important not to neglect the influences of accretion and migration on one another.
    The former cannot be understood without a better understanding of the second, and vice versa.
    The migration history, which results in varying accretion rates during the evolution, and the origin of accreted material, may be reflected in the planet's composition, and allow a better understanding of the late formation.
    This has to be studied in population synthesis models, which include more detailed models of the mutual effects of migration and accretion.
    Analytical models for planet growth such as \citet{tanigawa2007systematic} that do not include the planet migration, will underestimate the final masses of planets because they do not account for the gas crossing the gap and entering the Hill sphere.
    In agreement with previous studies \citep{nelson2000migration} we find that masses
of several Jupiter masses can be reached if single massive planets are allowed to migrate undisturbed through
their disks.

    Finally we note that in our simulations we modeled only 2D locally isothermal disks. It will be worthwhile in the
    future to extend these to full 3D.

    \begin{acknowledgements}
      We want to thank Aur\'{e}lien Crida and Bertram Bitsch for many useful discussions and helpful critisism.
      Some of the numerical simulations were performed on the bwGRiD cluster in T\"ubingen, which is funded by the
    state of Baden-W\"urttemberg,
    and the cluster of the Forschergruppe FOR 759 funded by the DFG.
    The work of Christoph D\"urmann was sponsored with a scholarship of the Cusanuswerk.
    \end{acknowledgements}

  \bibliographystyle{aa} 
  \bibliography{accretion} 
\end{document}